# Optical performance of prototype horn-coupled TES bolometer arrays for SAFARI

Michael D. Audley[*a], Gert de Lange[b], Jian-Rong Gao[b,c], Pourya Khosropanah[b], Richard Hijmering[b], Marcel L Ridder[b]

[a]SRON Netherlands Institute for Space Research, Postbus 800, 9700 AV Groningen, The Netherlands; [b]SRON Netherlands Institute for Space Research, Sorbonnelaan 2, 3584 CA Utrecht, The Netherlands; [c]Kavli Institute of Nanoscience, Delft University of Technology, Lorentzweg 1, 2628 CJ Delft, The Netherlands

**ABSTRACT**

The SAFARI Detector Test Facility is an ultra-low background optical testbed for characterizing ultra-sensitive prototype horn-coupled TES bolmeters for SAFARI, the grating spectrometer on board the proposed SPICA satellite. The testbed contains internal cold and hot black-body illuminators and a light-pipe for illumination with an external source. We have added reimaging optics to facilitate array optical measurements. The system is now being used for optical testing of prototype detector arrays read out with frequency-domain multiplexing. We present our latest optical measurements of prototype arrays and discuss these in terms of the instrument performance.

## 1. INTRODUCTION

The Japanese Space Agency's satellite observatory SPICA will use a large (~3-m diameter) primary mirror cooled to ≤8 K to enable high spectral-resolution, sky-background limited observations of the cold dusty Universe in the mid- and far-infrared[1]. The mission promises to revolutionize our knowledge of the origin and evolution of galaxies, stars and planetary systems. The SAFARI instrument was originally conceived as a far-infrared imaging Fourier transform spectrometer (FTS) for the SPICA satellite[2]. Since then, the design has been optimized for sensitivity[3] and SAFARI now uses diffraction gratings to provide a spectral resolution of ~300 and an in-line Martin-Puplett FTS to provide a high spectral-resolution mode. SAFARI will cover the spectral range λ=34—230 μm with four grating modules, each containing a diffraction grating and a linear detector array to sample the dispersed spectrum. The four detector arrays will contain a total of ~3300 Transition Edge Sensor[4] (TES) bolometers using Ti/Au superconducting bilayers on thin, thermally isolated silicon nitride islands. The bolometers sit behind feedhorns and in front of reflecting backshorts and incoming radiation is absorbed by a 7-nm thick Ta film.

SAFARI will allow us to study the dynamics and chemistry of a wide range of objects, including galaxies out to redshifts of $z \approx 5$—6. To take advantage of SPICA's low-background cold mirror, SAFARI's detectors require a dark noise equivalent power (NEP) less than $2 \times 10^{-19}$ W/√Hz. To achieve this sensitivity, the TES detectors have a transition temperature, $T_c$, of about 100 mK and are operated with a bath temperature of 50 mK.

Testing such sensitive detectors is challenging and requires careful attention to magnetic and RF shielding, stray-light exclusion, and vibration isolation. Both the TES and its SQUID readout are extremely sensitive to magnetic fields. Stray light exclusion is particularly important because TES detectors will saturate and become insensitive if the incident power is too high. SAFARI's detectors are over two orders of magnitude more sensitive than TES bolometers previously developed for ground-based applications[5] and they have correspondingly lower saturation powers (~5 fW).

We have built a test facility which will be used to qualify and characterize the SAFARI focal plane arrays and readout before they are integrated into the instrument. In addition to the strict requirements on background and interference, we require that this facility be flexible and re-configurable so that we can use it for dark and optical tests of single pixels through to the full focal-plane arrays. Through a systematic program of incremental modifications we improved the performance of the SAFARI Detector Test Facility to the point where it is now being used for routine measurements of prototype SAFARI detectors[6]. We have used it to confirm the high optical efficiency of prototype SAFARI single detectors and to understand their optical response in terms of the electromagnetic modes admitted by the feedhorns[7].

---

*audley@physics.org; phone +31 (0)50 363 9361; fax +31 (0)50 363 4033; www.sron.nl

## 2. DESCRIPTION OF THE TEST FACILITY

A Leiden Cryogenics dilution refrigerator[8] with a cooling power of ~200 μW at 100 mK forms the basis of he SAFARI Detector Test Facility (see Figure 1, left). We have found that we can reach a base temperature on the mixing chamber below 8 mK and have operated detectors in the dark with bath temperatures, $T_{bath}$, as low as 15 mK. This bath temperature is well below the value required for testing the SAFARI detectors ($T_c$~100 mK) and allows us to install experiments that place additional heat loads on the system, e.g. black-body illuminators. In practice, the parasitic heat loads from the various optical calibrators limit the detector bath temperature $T_{bath}$ to about 40 mK. The refrigerator is precooled by a Cryomech PT-415 pulse-tube cooler[9] which is attached to the 50-K and 3-K stages of the cryostat. To mitigate the effects of vibrations, the pulse-tube cooler has two expansion tanks and its rotary valve motor is separated from the cryostat. The expansion tanks and valve motor are mounted on the cryostat's support tripod. As well as requiring no expendable cryogens, the mechanical cooler is closer to the flight configuration which also uses mechanical cooling than a wet cryostat would be. It has the disadvantage, however, that we need to protect the detectors under test from mechanical and electrical interference from the cooler.

Wiring for detector readout and thermometry is provided by eight woven looms, each containing 12 twisted pairs. Two of these looms have Cu conductors for low electrical resistance (and hence low Johnson noise) and are used for the SQUID bias. The remaining looms have CuNi conductors to minimize thermal conductance. The looms are enclosed in stainless steel tubes and heatsunk at the various temperature stages of the cryostat. Stainless-steel and superconducting coaxial cables allow high-frequency measurements. RF shielding is provided by two nested Faraday cages. The outer one is formed by the Dewar main shell and contains the room-temperature readout electronics as well as a multiplexer box, which allows us to connect the room-temperature readout electronics to different SQUIDs. The inner Faraday cage is the 3-K shield. All wires entering the 3-K shield are low-pass filtered at several hundred MHz, with the exception of the superconducting coaxial cables. These are terminated in simple stub antennas and allow us to inject modulated GHz-range radiation in order to carry out RF-susceptibility measurements and to characterize the dynamic behavior of the bolometers[10]. A reconfigurable patch board on the 20-mK stage redistributes the signals from the looms between the two experiment boxes currently installed. Both of these experiment boxes are used for optical detector testing.

Each experiment box comprises a tin-plated copper can, with light-tight feedthroughs for wiring and an absorbing labyrinth where it attaches to its base, all surrounded by a Cryoperm can[10]. We have verified that this can provides good magnetic shielding and is light-tight. We had originally intended to replace the tin-plated copper can with a niobium can, but we found that the tin plating on the copper can provided adequate magnetic shielding. Due to its higher thermal conductivity, the copper can also cools more efficiently than a superconducting niobium shield would. The Cryoperm can provides a low-field environment for the tin-plated can in order to avoid trapping flux as the tin layer cools through its superconducting transition.

## 3. OPTIMIZATION OF THE TEST FACILITY

Our challenge was to build an ultra-low background test facility that still had the flexibility to accommodate the various experiments that will be needed until the SAFARI focal plane arrays have been fully characterized. Starting with a well-engineered system, we carried out a comprehensive program of optimization, making improvements to the system in small steps until we arrived at the point where were confident that the system was suitable for testing ultra-sensitive TES bolometers. The optimization program comprised many small improvements that came under three general areas.

### 3.1 Grounding and shielding

We paid careful attention to eliminating ground loops. We isolated the gas-handling system electrically from the cryostat and instrument rack. We made sure that the Faraday cages were closed by installing EMI/RFI shielding gaskets on all joints. The importance of this step may be seen by opening one of the boxes surrounding the room-temperature readout electronics: this increases the loading on the detectors by several fW.

### 3.2 Optimization of the pulse-tube cooler

We took some steps to mitigate the effect of vibrations and electrical interference from the pulse-tube cooler. We replaced the hose between the rotary valve and the cold head with a longer one that could better accommodate the motion of the valve and isolated it electrically from the cryostat. We also replaced the square-wave driver for the rotary-valve motor with a linear (sinusoidal) drive. This cleaned up the detector power spectra considerably, but introduced a strong 180-Hz mechanical resonance. We were able to eliminate this 180-Hz resonance by bending the hose between the rotary valve and the cold head into a loop.

### 3.3 Optimization of the readout wiring

We ensured that the signal wiring inside the cryostat was immobilized as much as possible. Because this system is a test bed and needs to be flexible, it was not always possible to use rigid conductors everywhere. The most critical wiring in the whole system is the circuit connecting the TES detector to the input coil of the SQUID. We made sure that any wiring associated with the input circuits of the SQUIDs was made from the stiffest wire available and shielded with superconducting Al tubing. This is the most important optimization we performed. Without it, the intrinsic noise of the TES was not even visible. With it, we could measure the detector NEP readily without any serious interference from the pulse-tube cooler.

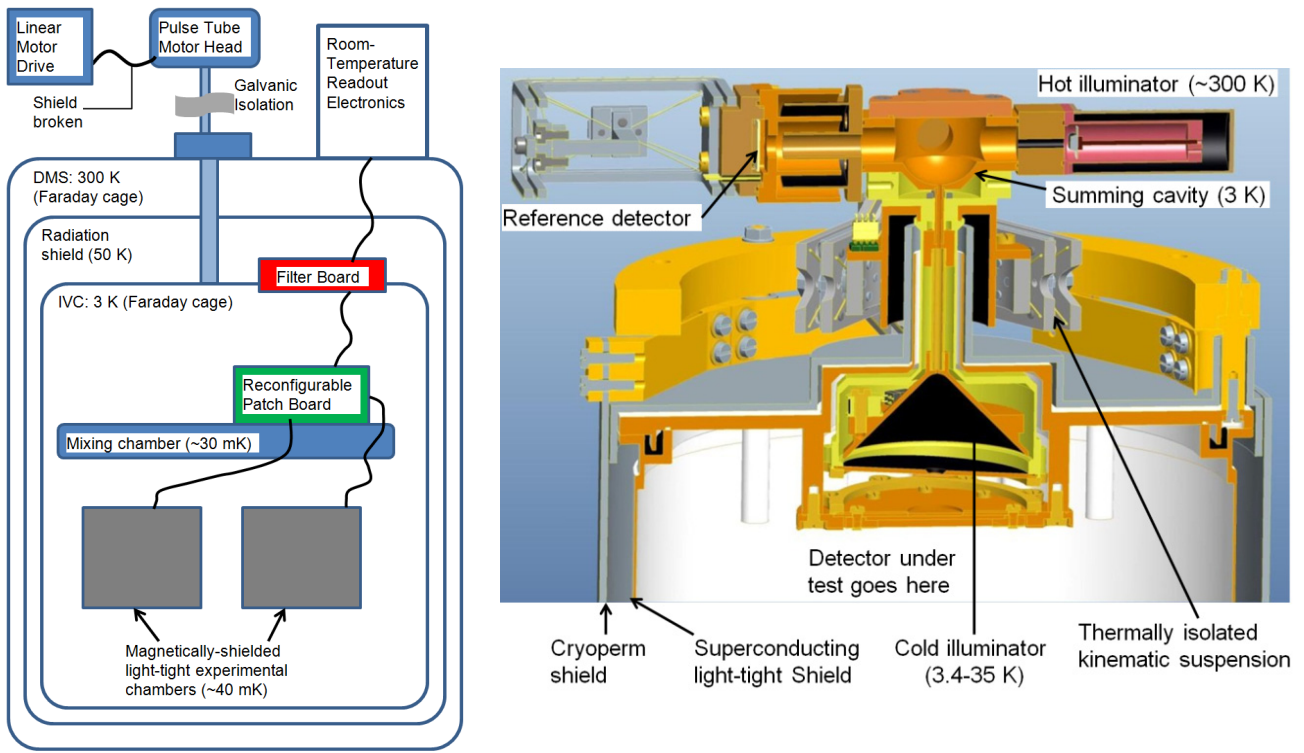

Figure 1 Left: Simplified schematic diagram of the cryostat. Right: Optical calibration source on magnetically shielded light-tight experimental chamber.

## 4. OPTICAL CALIBRATION SOURCE

We constructed a multi-functional optical calibration source using a reflective summing cavity to direct radiation to the detector under test (see Figure 1, right). The reflective summing cavity has four ports where radiation may be introduced from different calibration sources. It has a thermally isolated kinematic suspension and is heatsunk to the 3-K stage of the refrigerator.

### 4.1 Cold black-body illuminator

The cold black-body illuminator is a blackened cone whose temperature can be controlled in the range 3.4—34 K. This illuminator is surrounded by radiation shields heatsunk to the 3-K and 20-mK stages. Band-defining filters on the apertures of these shields allow only radiation around SAFARI's short-wave band to reach the detector. A light-pipe, with diameter 4.5 mm, emerges at a hole at the apex of this conical emitter, and allows radiation from the summing cavity also to illuminate the detectors under test. Thus, the detectors see a large, cold black-body source as well as a point source that emits radiation carried by the light-pipe from the summing cavity.

### 4.2 Hot black-body illuminator

We have successfully measured the optical response of SAFARI prototype detectors all the way to saturation using the cold illuminator[6],[7]. However, because of the low temperature (3.4—34 K) of this black-body emitter, the pass-band of the short-wave detectors is illuminated by the Wien tail of the black-body spectrum. This means that only the long-wavelength end of the pass-band sees significant power. In order to illuminate the pass-band more uniformly with a black-body source, we need that source to have a much higher temperature. We thus added a hot (200—300 K) source, with cold attenuation to reduce the power, to one of the ports of the summing cavity. Both this hot black-body illuminator and the cold illuminator described in Section 4.1 can be operated during the same cool-down, allowing cross-calibration.

### 4.3 Light-pipe

On another port of the summing cavity, there is a light-pipe (not to be confused with the short light-pipe described in Section 4.1) leading to room temperature so that we can inject power from an external FTS or a modulated source. This light-pipe can be closed with a double-bladed shutter on the 3-K stage. The shutter can be chopped at 20—40 Hz and has three stable positions. An attenuation factor on the order of $10^6$ is required to avoid saturating the detectors when the shutter is open. This attenuation is provided by cold (3 K) filters in the light-pipe and geometric dilution in the summing cavity. One of the blades of the shutter has apertures with three different diameters that allow us to adjust the amount of light entering the summing cavity from the light-pipe. Even with this flexibility, we expected that it might be difficult to achieve a suitable attenuation factor on the first try.

### 4.4 Reference detector

A reference detector is used to measure the spectral content of the injected radiation so that we can recover the spectral response of the detectors under test. The reference detector is a TES bolometer, suspended in a large absorbing cavity so that its optical absorption coefficient is flat over a broad wavelength range. We needed the reference detector to have a high saturation power in case the light-pipe attenuation factor was too low, but at the same time be as sensitive as possible in case it was too high. To increase the dynamic range of the reference detector we connected the Ti/Au bilayer in series with the Ta absorber, giving a double IV-curve[6]. We then had a detector that could operate in two regimes. For low incident powers, the Ta absorber would be superconducting and we would have a reasonably sensitive detector with an NEP of about 2 aW/√Hz. For high incident powers (>100 fW) the Ti/Au bilayer would be in the normal state and the Ta absorber would act as a TES with a parasitic resistance of 150 mΩ (the normal resistance of the Ti/Au bilayer) in series with it.

## 5. PERFORMANCE OF THE TEST FACILITY

In our initial tests of the optical calibration source we saw 500 fW of optical loading in the reference detector when the shutter was opened, but only 500 aW in the detector under test. The high power incident on the reference detector would have saturated it if we had not wired the Ta absorber in series with the Ti/Au bilayer. However, it was clear that adjustment was needed. We therefore added a mirror to the summing chamber (see Figure 2, left). This directs more light from the light-pipe to the detector under test and reduces the power to the reference detector. Originally, this mirror was flat, but it has since been replaced by an ellipsoidal mirror that gives better coupling between the light-pipe and the detector under test. A groove cut in the mirror block prevents the hot black-body source and reference detector from being completely blocked off. In this configuration we see about 60 fW in the reference detector and 3.9 fW in the detector under test when the shutter is opened. Figure 3 shows the photon noise in the detector under test when the light-pipe is illuminated with a chopped glowbar (a 1200-K black-body source). When the shutter is opened we see the 40-Hz chopped signal sitting more than 20 dB above the 300-K photon noise. The low-frequency noise in the power spectrum is intrinsic to the SQUID and the difference in roll-off at high frequency is caused by different amounts of ADC noise being added by different preamplifier gains. This means that the performance of the SAFARI Detector Test Facility is dominated by the readout and not the refrigerator. We have also used a prototype SAFARI detector to measure interferograms from an external FTS through the light-pipe with high contrast. The spectra derived from these interferograms are consistent with the transmission of the filter stack[12].

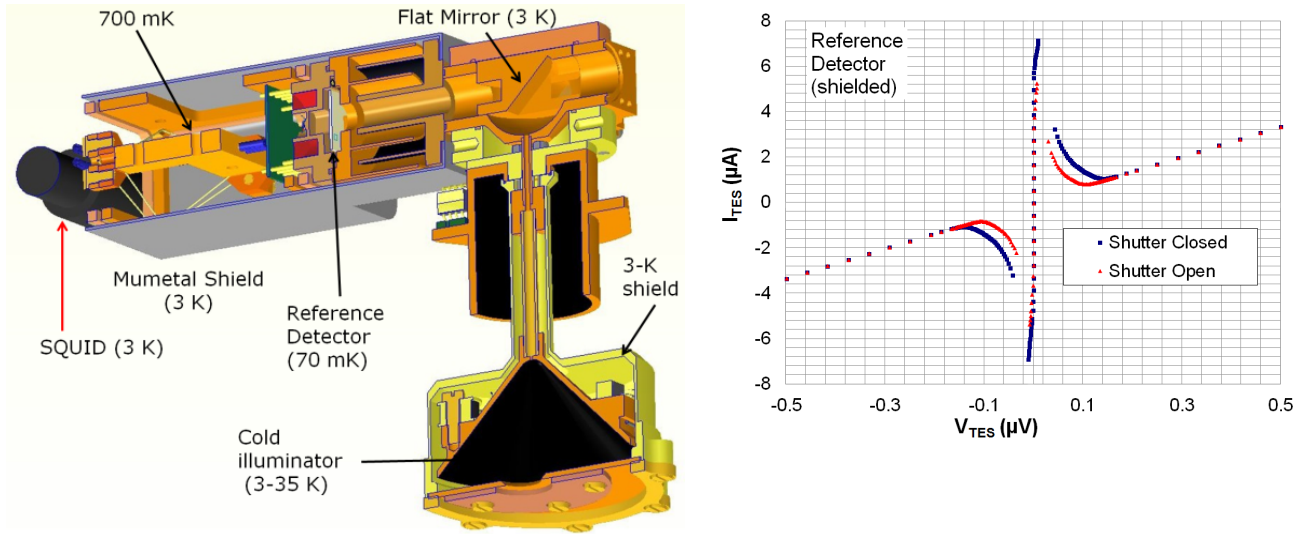

Figure 2 Left: Optical calibration source with a mirror to increase the throughput from the light-pipe to the detector under test. Right: TES IV curves for the reference detector with the shutter open and closed; the optical loading is 60 fW when the shutter is opened.

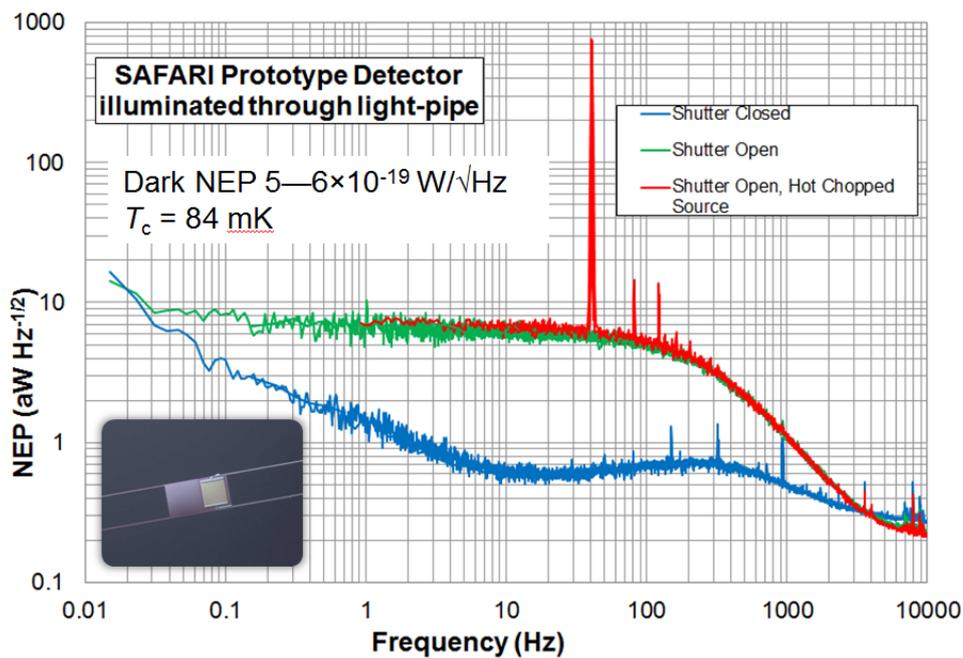

Figure 3 Photon noise measured in the ultra-sensitive prototype SAFARI TES bolometer shown in the inset photograph. The detector was illuminated through the light-pipe with an external hot source chopped at 40 Hz. The lowest blue curve is the detector power noise when the light-pipe's shutter is closed. The green curve is the noise when the shutter is opened and the detector sees an optical load of 3.9 fW. The red curve shows the noise in the detector with the chopped hot external load, which adds a high-contrast peak at 40 Hz. This is the most sensitive detector measured optically in the 30—60-μm range.

# 6. OPTICAL MEASUREMENTS OF PROTOTYPE ARRAY

We mounted a 72-pixel array of prototype detectors with NEP~2×10-18 W/√Hz in the test facility. These detectors were designed for SAFARI's short-wave band, with the wavelength range λ=34—60 μm. Each detector sat above a hemispherical backshort[6]. In front of the array was an 8×8 array of electroformed rectangular feedhorns, with a pixel pitch of 830 μm, as shown in Figure 4. The detectors were AC-biased at frequencies between 1.1 and 3.4 MHz and read out using baseband feedback[8],[14]. Since this was a very early prototype array, not all pixels were connected. However, all 38 pixels that were connected were found to work.

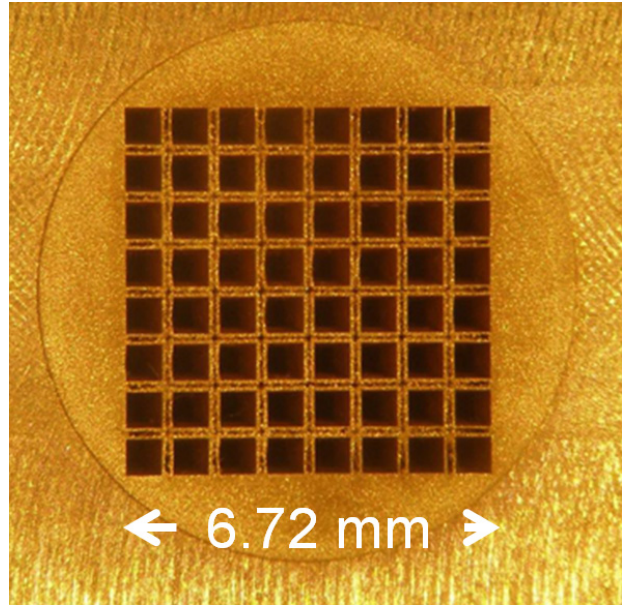

Figure 4 Photograph of the 8x8 feedhorn array. The pixel pitch is 830 μm. This feedhorn array is mounted in front of the detector array

We measured IV curves of the 38 connected detectors with the black-body illuminator at its base temperature (3 K) and also at 22 K. At the higher illuminator temperature some of the Joule power dissipated in the detectors was replaced by absorbed optical power, so that the 22-K power plateaus were depressed relative to the 3-K power plateaus. We took the absorbed optical power to be the difference in the Joule power measured at the point on the TES transition where the TES resistance is 80% of the normal resistance. An example of dark and optical power plateaus is shown in Figure 5. The results for the entire array are shown in Figure 6 and the histogram in Figure 7 shows the distribution of detected optical powers. We have normalized the optical power to the saturation power of each TES to allow comparison between pixels. This was necessary because we are using a very early generation of LC filters that causes an apparent decrease in power plateaus as the bias frequency increases. This effect is absent in later generations of the design. The detector array itself appears to be remarkably uniform. Every working pixel detected significant optical power, while no pixel was saturated. As can be seen in Figure 7, the every pixel detected between 15% and 55% of its saturation power.

We have now installed reimaging optics in the test facility, as shown in Figure 8. The mirror is interchangeable and we have fabricated flat, ellipsoidal, and parabolic mirrors for different experiments. Measurements with an ellipsoidal mirror are ongoing.

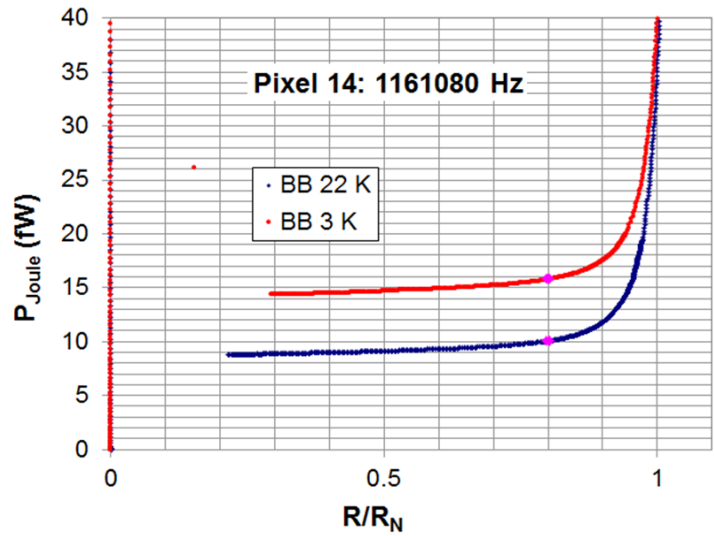

Figure 5 Measured power plateaus for one of the pixels in the array with the cold black-body illuminator at 3 K and at 22 K. The absorbed optical power is derived from the difference in the Joule power at $R/R_N = 0.8$. The 22-K power plateau is depressed relative to the 3-K power plateau dur to the absorption of optical power.

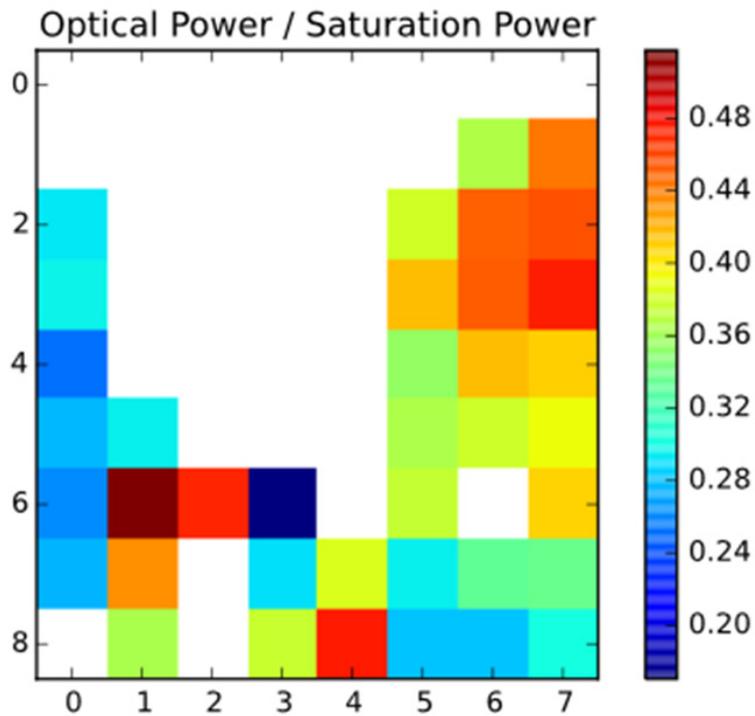

Figure 6 Optical power detected by array from the cold black-body illuminator at a temperature of 22 K. Each of the 38 connected pixels is shown with a color corresponding to the optical power it detected. Pixels that are not connected are shown as white. The detected optical power has been normalized to the TES saturation power. Note that every connected pixel detected significant optical power and no pixel was saturated.

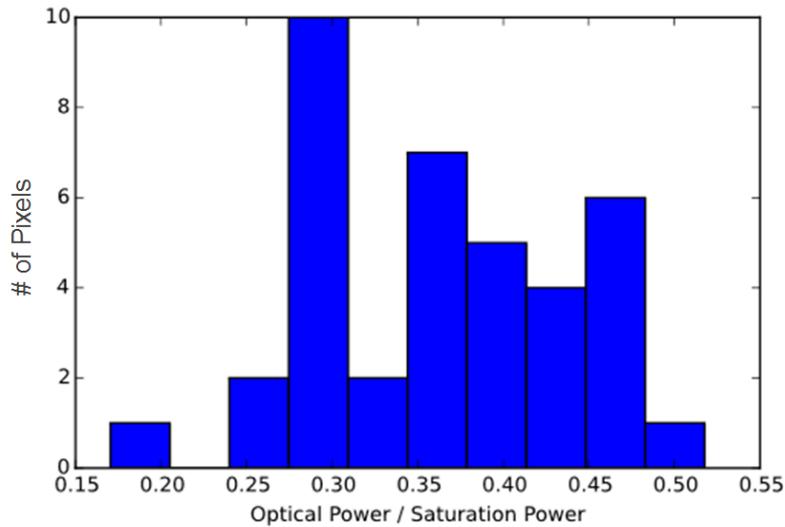

Figure 7 Number of pixels detecting optical power in different ranges from the 22-K black-body illuminator. Note the narrow range of detected powers. Every connected pixel detected significant optical power and no pixel was saturated.

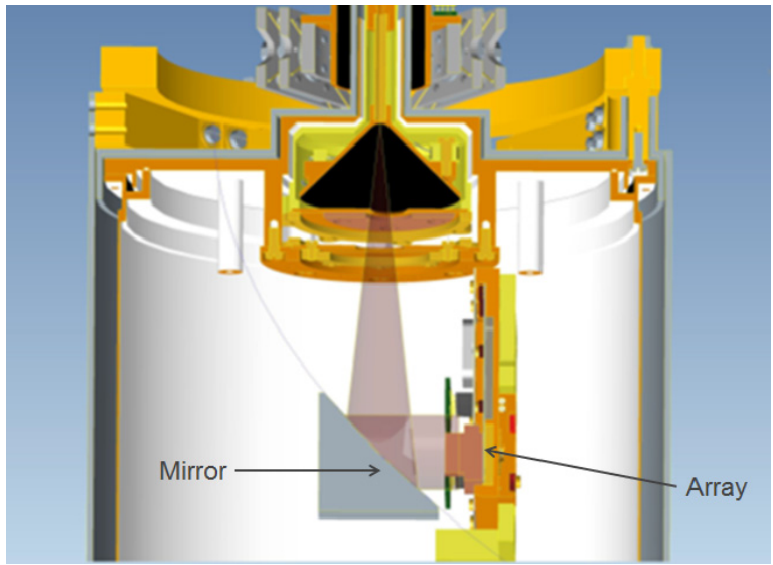

Figure 8 Reimaging optics for array measurements. Flat, ellipsoidal, or parabolic mirrors can be installed, depending on the desired measurement.

## 7. CONCLUSIONS

We have built up a low-background millikelvin test facility suitable for optical characterization of ultra-low-NEP far-infrared detectors. Our preliminary measurements demonstrate that an array of prototype SAFARI detectors detects light with high optical efficiency and good uniformity. This is important for the SAFARI instrument, which requires an optical efficiency of 50%, in combination with the NEP of $<2\times10^{-19}$ W/$\sqrt{Hz}$. Our ongoing measurements with reimaging mirrors will allow us to measure optical crosstalk, which is crucial for recovering the dispersed grating spectra.